\documentclass[final,5p,times,twocolumn]{elsarticle}
\usepackage{graphicx}
\usepackage{amssymb}
\usepackage{lineno}
\usepackage{color}
\usepackage{natbib}

\begin{document}

\newcommand{\red}{\textcolor{red}}
\newcommand{\etc}{\textit{etc.}}
\newcommand{\ie}{\textit{i.e.\,}}
\newcommand{\eg}{\textit{e.g.\,}}
\newcommand{\egcomma}{\textit{e.g.,\,}}
\newcommand{\etal}{\textit{et al.\,}}
\newcommand{\etalcomma}{\textit{et al.,\,}}
\newcommand{\roha}{Rohacell\textsuperscript{\textregistered} }
\newcommand{\rohacomma}{Rohacell\textsuperscript{\textregistered}, }
\newcommand{\Tedlar}{Tedlar\textsuperscript{\textregistered} }

\begin{frontmatter}

\title{GEANT4 Simulation of a Scintillating-Fibre Tracker for the Cosmic-ray Muon Tomography of Legacy Nuclear Waste Containers}

\author[Glasgow]{A.\,Clarkson}
\author[Glasgow]{D.\,J.\,Hamilton}
\author[Glasgow]{M.\,Hoek\fnref{Curr}}
\author[Glasgow]{D.\,G.\,Ireland}
\author[NNL]{J.\,R.\,Johnstone}
\author[Glasgow]{R.\,Kaiser}
\author[Glasgow]{T.\,Keri}
\author[Glasgow]{S.\,Lumsden}
\author[Glasgow]{D.\,F.\,Mahon}
\author[Glasgow]{B.\,McKinnon}
\author[Glasgow]{M.\,Murray}
\author[Glasgow]{S.\,Nutbeam-Tuffs}
\author[NNL]{C.\,Shearer}
\author[NNL]{C.\,Staines}
\author[Glasgow]{G.\,Yang}
\author[NNL]{C.\,Zimmerman}

\address[Glasgow]{SUPA, School Of Physics \& Astronomy, University of Glasgow, Kelvin Building, University Avenue, Glasgow, G12 8QQ, Scotland, UK}
\address[NNL]{National Nuclear Laboratory, Central Laboratory, Sellafield, Seascale, Cumbria, CA20 1PG, England, UK}
\fntext[Curr]{Current affiliation:  Johannes Gutenberg-Universit\"{a}t, Mainz}

\begin{abstract}
Cosmic-ray muons are highly penetrative charged particles that are observed at sea level with a flux of approximately one per square centimetre per minute.  They interact with matter primarily through Coulomb scattering, which is exploited in the field of muon tomography to image shielded objects in a wide range of applications.   In this paper, simulation studies are presented that assess the feasibility of a scintillating-fibre tracker system for use in the identification and characterisation of nuclear materials stored within industrial legacy waste containers.  A system consisting of a pair of tracking modules above and a pair below the volume to be assayed is simulated within the GEANT4 framework using a range of potential fibre pitches and module separations.  Each module comprises two orthogonal planes of fibres that allow the reconstruction of the initial and Coulomb-scattered muon trajectories.  A likelihood-based image reconstruction algorithm has been developed that allows the container content to be determined with respect to the atomic number $Z$ of the scattering material.  Images reconstructed from this simulation are presented for a range of anticipated scenarios that highlight the expected image resolution and the potential of this system for the identification of high-$Z$ materials within a shielded, concrete-filled container.  First results from a constructed prototype system are presented in comparison with those from a detailed simulation.  Excellent agreement between experimental data and simulation is observed showing clear discrimination between the different materials assayed throughout.
\end{abstract}


\end{frontmatter}


\section{Introduction}
Cosmic-ray muons occur in nature from the decay of high-energy pions produced in the Earth's upper atmosphere.  These fundamental particles have similar properties to electrons, though are several hundred times more massive, making them much more penetrative.  As a consequence of this, they are observed at sea level with a flux of approximately one muon per square centimetre per minute with momenta of several GeV\,c$^{-1}$.   As charged particles, they interact weakly with matter via ionising interactions with atomic electrons and Coulomb scattering off nuclei.  

These properties are increasingly being exploited in the growing field of Muon Tomography (MT) to assay the internal structure of shielded objects that cannot be interrogated using conventional forms of imaging radiation such as X-rays.  Since their first reported use in radiography by E.\,P.\,George~\citep{George1955} in the 1950s, there has been a wealth of applications exploiting cosmic-ray muons for imaging purposes, most notably L.\,W.\,Alvarez's search for hidden chambers within the Second Pyramid of Chephren in Egypt~\citep{Alvarez1970} in the late 1960s.  Recently, there has been a significant, concerted increase in the exploitation of MT for volcanic assaying~\citep{Tanaka2005,Ambrosi2011} and also for improving national security~\cite{Borozdin2003a,Gnanvo08} with the development of cargo scanners for nuclear threat detection.

Before the turn of the century, the assaying of large and/or dense structures utilised the attenuation properties of muons, characterised by the attenuation length $X_{0}$ of the passive material.  Seminal work presented by Borozdin \etal\,in Ref.~\citep{Borozdin2003a} introduced the potential of using the Coulomb scattering of muons for the identification of potentially-hazardous high-$Z$ materials concealed within (comparatively) small, shielded containers.  In this approach, the identification of the initial and Coulomb-scattered muon trajectories allows the scattering density $\lambda$ and location to be determined.  Here, $\lambda$ which is related to $X_{0}$, is known to exhibit an inherent dependence on the atomic number $Z$ of the scattering material~\citep{Schultz04}.

\begin{figure*}[t] 
\centering 
\includegraphics[width=2.0\columnwidth,keepaspectratio]{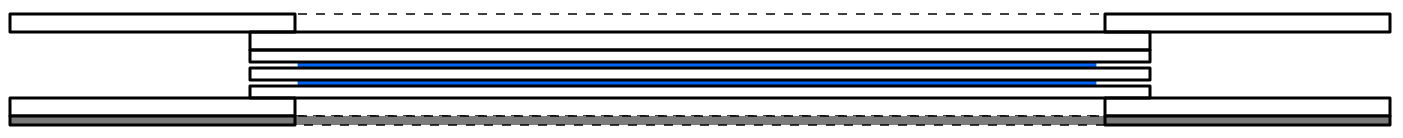}
\caption{Schematic of the simulated module composition.  Shown is a cross-section through the centre of the module highlighting the scintillating fibres (blue), \roha support sheets (white) and the aluminium base plate (grey).  The dashed lines represent the removed sections of \roha and aluminium within the active volume.  Top-down, \roha thicknesses of 6\,mm, 6\,mm, 4\,mm, 6\,mm, 4\,mm and 6\,mm support the fibre layers and provide rigidity in the constructed module structure.  Not simulated are the vertical support frame and other mechanical support structures which are outwith the active volume.  The thin lightproof covering is not simulated.}
\label{fig:module}
\end{figure*}

This work presents the preparatory studies performed using the high-energy physics simulation framework GEANT4~\citep{geant4}, which influenced the design and construction of a prototype MT system~\cite{VCI}.  These studies assessed the feasibility of using a system based on a round, plastic scintillating fibre detection medium for the non-destructive assay of potential nuclear materials stored within highly-engineered legacy waste containers.  For this application, clear discrimination between low, medium and high-$Z$ materials within a concrete-filled container was essential along with accurate position determination of any identified objects.  

The GEANT4 simulation of the prototype detector system, and the muon event generator used are outlined in Section~\ref{sec:sim} and the probabilistic image reconstruction method used throughout this work is detailed in Section~\ref{sec:MLEM}.   Simulation studies, shown in Section~\ref{results}, were performed to assess various aspects required for a constructed prototype system including the fibre pitch and module positions.  The anticipated image resolution and degree of material discrimination expected from the system are also outlined.   A test configuration of materials was imaged using a prototype system constructed based on these results and is compared with simulated data in Section~\ref{sec:TestSetup}.  Images reconstructed from a detailed GEANT4 simulation of a small-scale waste container are presented in Section~\ref{sec:canister}.  In the future, it is foreseen that this system will be deployed within the UK nuclear industry to assay legacy nuclear waste containers.

\section{GEANT4 Simulation of the Prototype MT System}\label{sec:sim}
Prior to the fabrication of a small-scale prototype MT detector, dedicated studies were performed using a detailed simulation of the proposed setup using GEANT4.  These studies were designed to establish key specifications for the final system such as the optimal fibre pitch and detector positions, as well as providing an indication of the expected data collection durations required to obtain the degree of material discrimination and image resolution necessary for this application.
   
This simulated system was comprised of four tracker modules, two situated above and two below the volume under interrogation (referred to as the assay volume).  Each module consisted of two orthogonal detection planes, each comprising a single layer of round plastic scintillating fibres.  The identification of the two struck fibres per module yields a single space point.  A space point in each of the four modules allows the reconstruction of the initial and Coulomb-scattered muon trajectories that are required to reconstruct an image.  

\subsection{Module Composition}
A schematic of the simulated module configuration is shown in Figure~\ref{fig:module}.  The chosen design consisted of two orthogonal layers of round scintillating fibres with a polystyrene-based core (97\% by cross-sectional diameter) and polymethylmethacrylate (PMMA) cladding (3\%).  These were supported by various thicknesses of \footnote{http://www.rohacell.com}\rohacomma a low-density rigid foam structure and an aluminium base plate.   The dimensions of the active area per module provided by the fibres remained fixed at 256\,mm by 256\,mm throughout the studies reported in this work.    Fibre pitch studies which influenced the choice of fibre used in the constructed prototype are presented in Section~\ref{sub:pitch}.     In total, the active area was surrounded by 20\,mm of \roha to provide sufficient support in the constructed system.  Simulation studies showed that this additional support material had a negligible effect on the Coulomb scattering of the simulated muons, and hence on the reconstructed images.  Two additional layers of \rohacomma both measuring 6\,mm in thickness, and the 3\,mm-thick aluminium baseplate used to secure the module to a vertical support stand were also simulated.  These are shown at the top and bottom of the schematic in Figure~\ref{fig:module} and do not contribute to the material in the active volume due to a removed inner area of 270\,mm\,x\,270\,mm.  All other components required for the constructed prototype system, \eg vertical support frame, PMTs, additional support mechanisms, thin lightproof coverings \etc, have not been simulated as these were either sufficiently far from the active volume and/or have been shown to provide a negligible impact on the passage of muons through the system. 

\subsection{Detector Configuration}
Four of the modules described in the previous subsection were simulated in perfect alignment, parallel, and in the same vertical plane.  The inter-modular separations for the top and bottom module pairs, along with the assay volume dimension, were chosen based on the studies reported in Section~\ref{sub:spacings}.


\subsection{Muon Event Generator}\label{geant4}
\begin{figure}[t] 
\centering 
\includegraphics[width=\columnwidth,keepaspectratio]{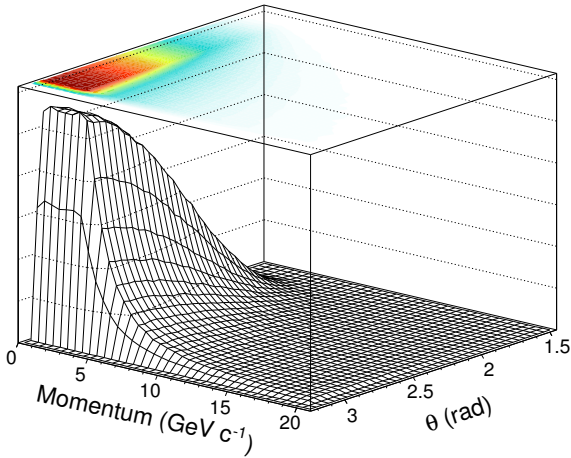}
\caption{Monte Carlo simulated momenta and $\theta$ angle distributions for the cosmic-ray muon event generator described in the text.  These distributions are propagated through the GEANT4 simulations of the prototype system described in Section~\ref{results} to allow the determination of the expected image quality and corresponding exposure times for future experimental results.  Here, a value of $\pi$\,rad corresponds to a muon trajectory which is vertically down in this $\theta$ convention.}
\label{fig:muonMC}
\end{figure}

A standalone Monte Carlo event generator was developed for the purposes of the simulation studies presented in this work.  This generator implemented well-defined characteristics of cosmic-ray muons including a sea-level flux of approximately 1\,cm$^{-2}$\,min$^{-1}$, a flat momentum distribution up to the mean value $p_{\mathrm{mean}}$ of 3.35\,GeV\,c$^{-1}$ with a $p^{-2.7}$ distribution at higher momenta, and a $\cos^{2}\theta$ angular dependence~\cite{PDG}.  Muons were uniformly generated in a plane, parallel to the detector modules, above the detector system.  The momentum and angular properties are shown in Figure~\ref{fig:muonMC} for a randomly generated sample of muons.  Throughout this work, this muonic distribution was propagated through the various detector simulations within the GEANT4 framework which incorporated all necessary muonic interactions, and their relative probabilities of occurring, with matter. 

\section{Image Reconstruction}\label{sec:MLEM}
\subsection{Maximum Likelihood Expectation Maximisation}
An iterative image reconstruction algorithm based on the probabilistic Maximum Likelihood Expectation Maximisation (MLEM) method, introduced in Ref.~\cite{Schultz07} by Schultz \etalcomma was further developed for this application using detailed simulations of the prototype detector setup described previously.   Prior to performing the imaging analysis, the assay volume was divided into small volume elements called voxels.  The dimensions of these voxels were pre-determined by the analyser and influenced the necessary data collection duration and achievable image resolution in the x and y directions, \ie smaller voxels provide greater definition in the final images but require a longer collection period due to fewer muons passing through their smaller volume.  

In the analysis, the incoming and Coulomb-scattered vectors of every muon that passed through the entire acceptance of the MT system, \ie depositing signals in all eight layers, were projected to their Point of Closest Approach (PoCA).  Provided that these two vectors were within a pre-set tolerance (referred to as the Distance of Closest Approach or DoCA) of each other at this position, the MLEM algorithm determined a normalised probability of scattering in each voxel that the muon was considered to have passed through.  This probability was weighted in each voxel by several factors including the pathlength of the muon within that element.  

\begin{figure}[t] 
\centering 
\includegraphics[width=\columnwidth,keepaspectratio]{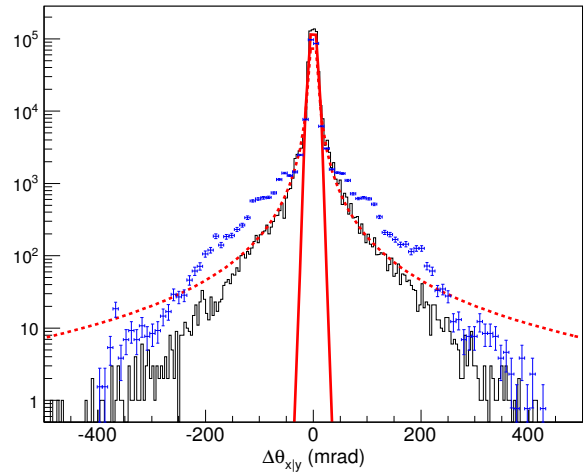}
\caption{Scattering angle distribution from a sample of simulated (black line) and experimental (blue points) data highlighting the differences between Cauchy-Lorentzian (dashed red line) and Gaussian (solid red line) fits.   The relative lack of simulated data at larger scattering angles, \ie $\left|\Delta\theta_{x|y}\right| >$ 200\,mrad, is a direct result of muons scattering outwith the finite acceptance of the detector system.  All data is shown prior to DoCA restrictions.}
\label{fig:GausvCauchy}
\end{figure}

With no scattering information determinable within the assay volume, only the average spatial and angular deviations experienced by the Coulomb-scattered muon were reconstructed.  The algorithm assumed all muons to have undergone a single, significant scatter at the PoCA position.  As such, potential discrepancies may arise from muons which have scattered significantly multiple times as they passed through the volume.    The scattering distribution of cosmic-ray muons is Cauchy-Lorentzian in nature, though in the MLEM algorithm, the scattering density $\lambda$ is related to the covariance matrix of a Gaussian model which accurately describes the low-scattering region that accounts for approximately 95\% of data.   A comparison between Cauchy-Lorentzian and Gaussian fits to simulated scattering data (along with experimental data, described later in Section~\ref{sec:TestSetup}) is shown in Figure~\ref{fig:GausvCauchy}, with the former yielding a FWHM value of 7.5\,mrad.   The discrepancies between data and simulation at larger angles result from misidentified events which are mostly removed prior to MLEM analysis   Here, the tails, which are not described by the Gaussian model and thus have the potential to skew the reconstructed images, are also observed at larger scattering angles. 

After many muons had passed through the system, the most likely scattering density $\lambda$ in each voxel was calculated via an iterative procedure which began with an initial $\lambda$ value pre-assigned in each voxel.  This was conventionally set to the expected value of the predominant material within the system \ie air or concrete in this work.   This ensured quicker convergence of the $\lambda$ values, which were used as the imaging metric in all the results presented in the following sections.  

The assay volume was chosen as a cube of dimension 300\,mm in the central region of the detector assembly with cubic voxels of 10\,mm dimension used throughout this work unless otherwise stated.   Each voxel was assigned a unique hexadecimal identifier which allowed it to be populated with scattering information during imaging analysis.

\begin{figure*}[t] 
\centering 
\includegraphics[width=0.66\columnwidth,keepaspectratio]{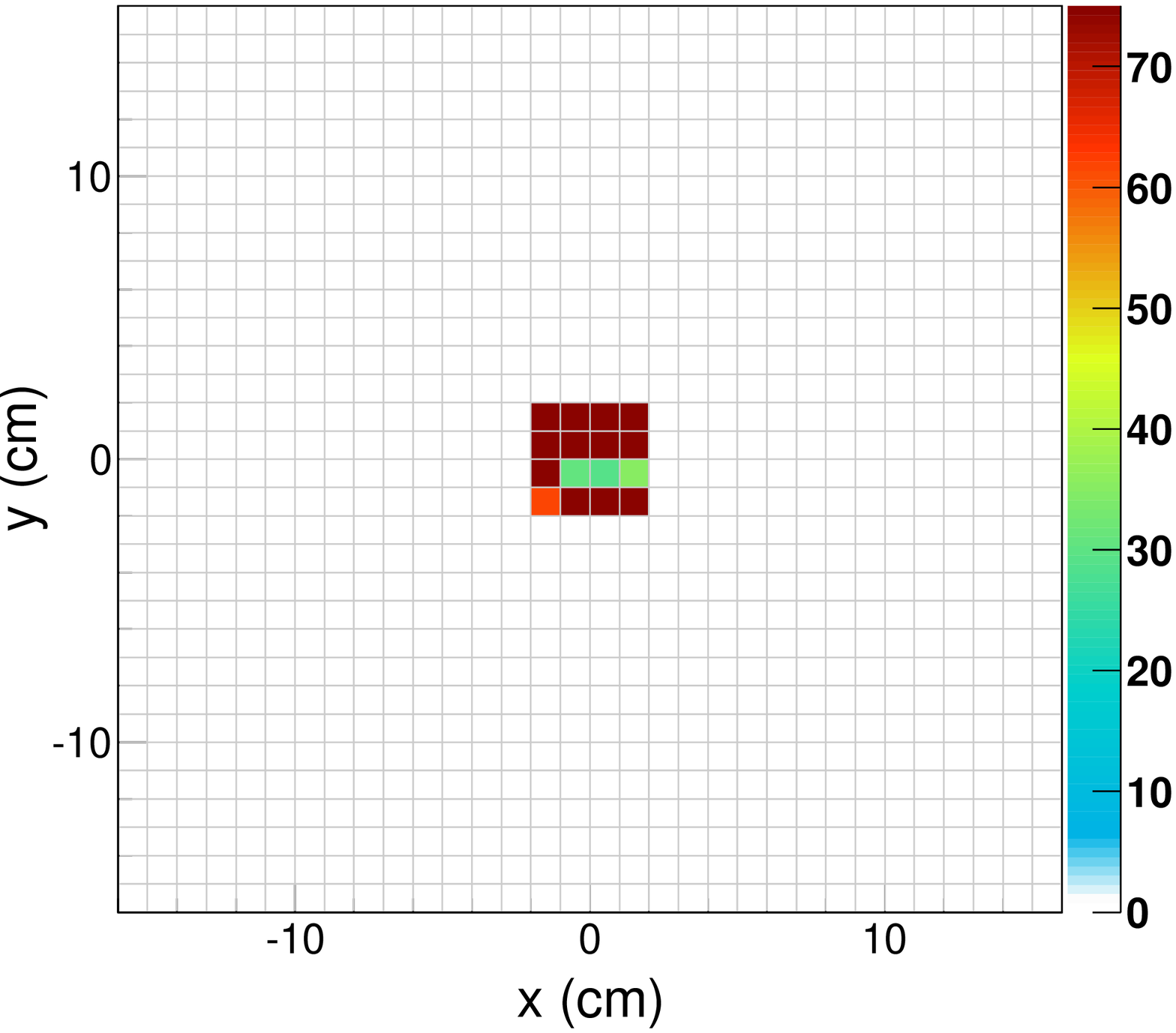}
\includegraphics[width=0.66\columnwidth,keepaspectratio]{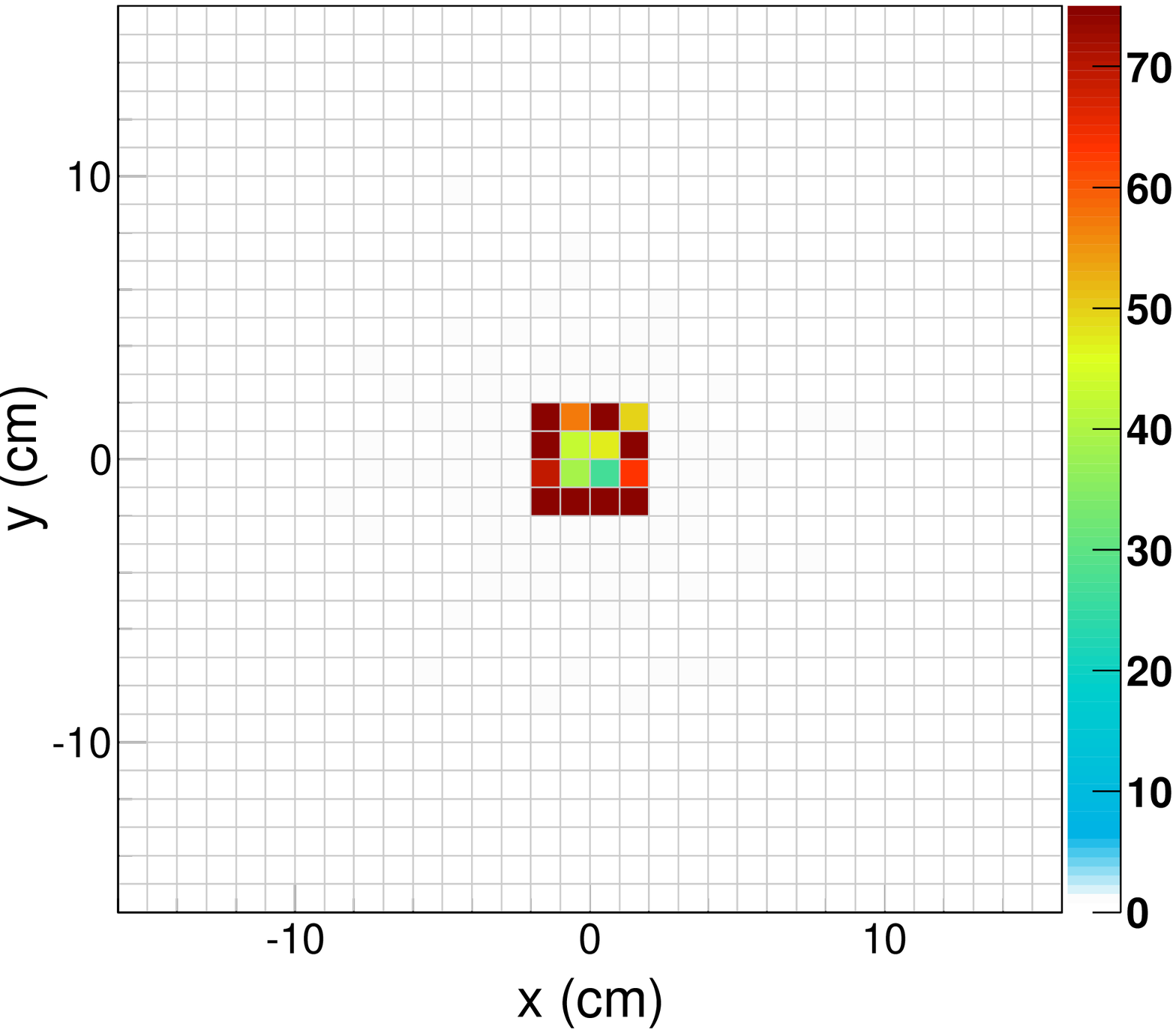}
\includegraphics[width=0.66\columnwidth,keepaspectratio]{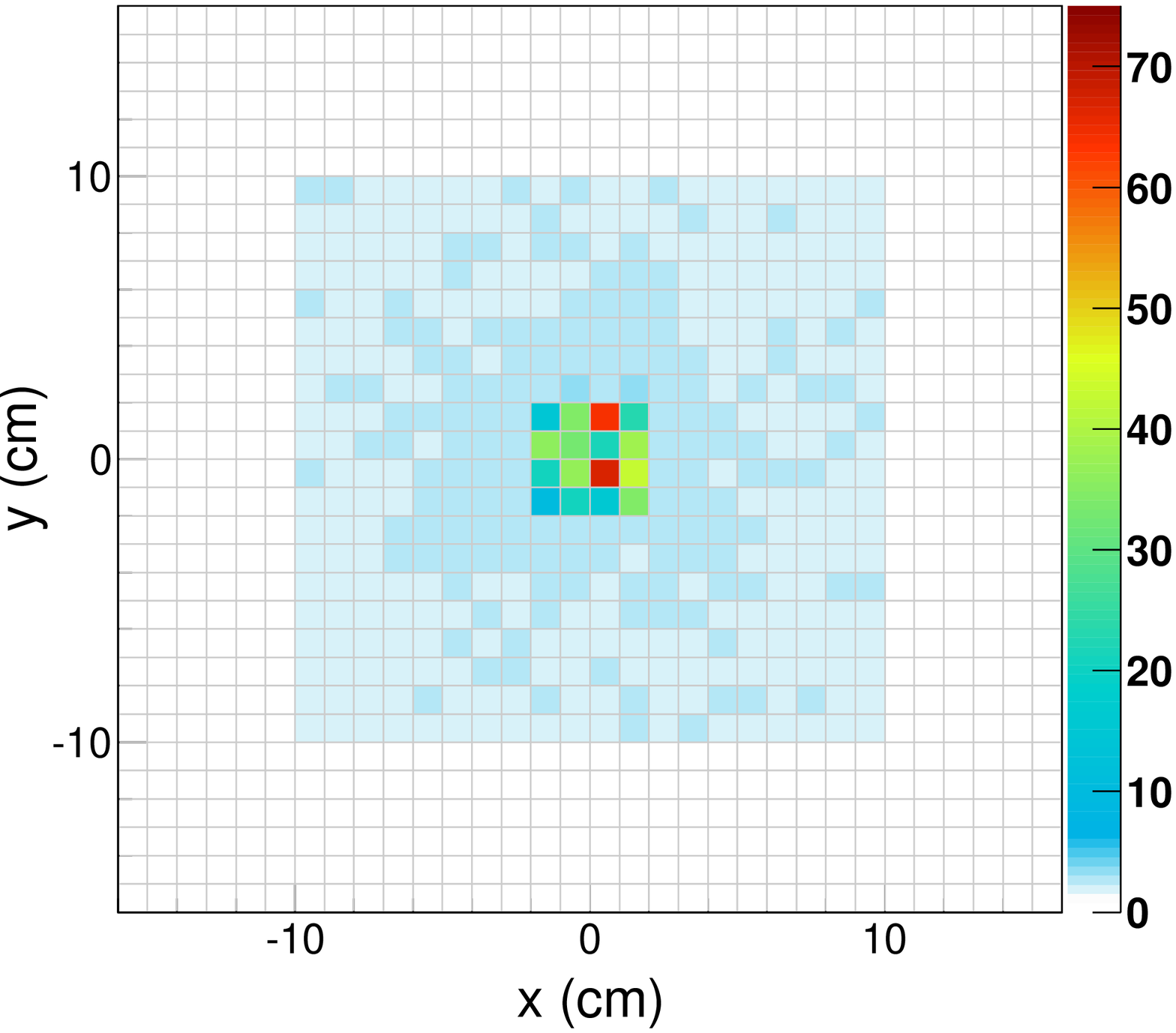}
\caption{Comparison between the reconstructed images in the central 10\,mm slice in the z plane for 1\,mm (left), 2\,mm (centre) and 4\,mm (right) fibre pitches.  The colour scale indicates the most-likely $\lambda$ values within each voxel reconstructed from by the MLEM algorithm.  Voxel boundaries are indicated by the light grey lines.  Similar results are observed in the two smaller pitch cases with the uranium clearly visible within the matrix.  The 4\,mm case does not reconstruct the uranium as clearly, though the 16 occupied voxels are still discriminated from the surrounding air.  In all three scenarios, the excellent image resolution in the x and y directions is noted. }
\label{fig:PitchXY}
\end{figure*}

\subsection{Event Selection Criteria}
Once the generated sample of muons was propagated though the GEANT4 simulation of the scintillating fibre detector system, the unique indices of the eight struck fibres were recorded for each event.   This provided an overall detection efficiency of 78\%, corresponding to the maximum efficiency per layer of 97\% provided by the cross-sectional width of the active core of the fibre.  The reasonable assumption that the constructed prototype could identify this muon to a high degree of reliability was made, albeit in reality this would be achieved with reduced detection efficiency due to light transmission losses within the fibres and potential dead channels arising from the construction process.

Events which registered a hit in each of the eight detection layers were considered as candidates for imaging analysis.  Each candidate event was required to have a PoCA position within the assay volume and a DoCA restriction of 10\,mm was imposed on the reconstructed vectors to remove potentially erroneous events or those with several significant scatters that the algorithm would fail to reliably evaluate.  The magnitude of the scattering angle components along the x and y directions, denoted $\Delta\theta_x$ and $\Delta\theta_y$, were limited to 100\,mrad to minimise the effects of the Cauchy-Lorentzian tails within the data.  These had the potential effect of producing large scattering (and hence large $\lambda$ values) within the reconstructed images that had originated from low-$Z$ materials.  An event was not considered if no detectable scattering had occurred.  Finally, no momenta restrictions were imposed upon the data.  Studies into the influence of individual muon momenta are discussed in Section~\ref{sub:momenta}.

\subsection{Imaging Parameters} 
Unless otherwise stated, all images have been reconstructed using a substantial amount of simulated data (amounting to several months of muon exposure for this small acceptance prototype), and have been iterated until the $\lambda$ values had converged in the majority of the voxels.  The number of iterations required may not be constant in cases of different simulated geometries and materials.   In all cases, the choice of the initial $\lambda$ values per voxel was representative of the matrix used \ie air or concrete.  All error matrices, which account for the propagated effects on the scattering angles and spatial deviations from the intrinsic resolution of the fibres, have been determined from each analysed data set using the conventions outlined in Refs.~\cite{Schultz07,SchultzThesis}.  Where several scenarios are being compared, the same generated muon event sample has been used for each.  All other criteria have been introduced in the previous sections.

\section{GEANT4 Simulation Studies}\label{results}
Dedicated simulation studies were performed to assess various aspects of the detector system prior to construction, some of which have been mentioned in the previous sections.  All studies were devised to influence the design and fabrication processes and to optimise the quality of the images reconstructed by this system.

\subsection{Fibre Pitch Requirements}\label{sub:pitch}

\begin{figure}[t] 
\centering 
\includegraphics[width=\columnwidth,keepaspectratio]{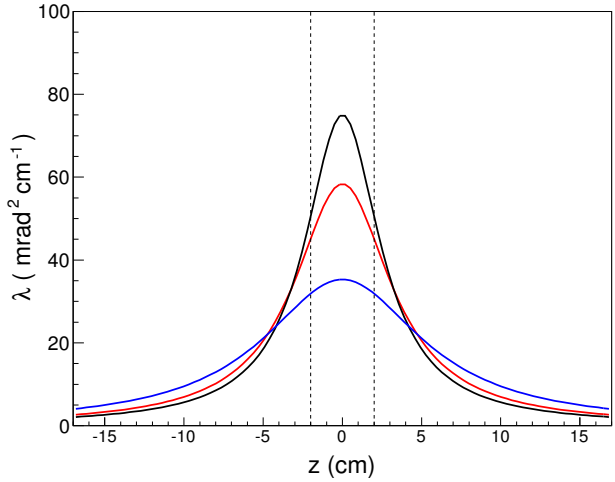}
\caption{Comparison between the z-direction resolution observed in the reconstructed images using 1\,mm (black), 2\,mm (red) and 4\,mm (blue) fibre pitches.  Shown are Cauchy-Lorentzian fits to the reconstructed $\lambda$ distribution in the z-direction.  An average over all slices containing the uranium was taken.  The mean values of all three fits showed slight deviations from zero.  For comparison, these have been centred on zero.  The dashed lines highlight the dimension of the simulated block.}
\label{fig:fibrepitch}
\end{figure}

Prior to the construction of the small-scale prototype detector system, studies were performed to assess the fibre pitch requirements.  Three pitches of \footnote{http://www.saint-gobain.com}Saint-Gobain BCF-10 round plastic scintillating fibres, all with cladding of 3\% cross-sectional diameter, were investigated: 1\,mm, 2\,mm and 4\,mm.  All were shown to provide sufficient light transmission for interaction with cosmic-ray muons and to emit light within the wavelength range provided by the chosen photon detector, the \footnote{http://www.hamamatsu.com}Hamamatsu H8500 MAPMT.  A simple test case with detector geometry of equal dimensions was simulated for each of the 1\,mm, 2\,mm and 4\,mm pitch sizes.  These correspond to 256, 128 and 64 fibres respectively per detection layer.  For this study, a single 40\,mm cube of uranium was simulated at the centre of the assay volume.  In all three cases, this cube was resolved in the x and y directions to roughly similar degrees of precision - see Figure~\ref{fig:PitchXY}.   However, the average $\lambda$ value reconstructed using the 4\,mm pitch was approximately 50\% lower than in the other two scenarios.

The reconstructed resolutions in the z direction are shown in Figure~\ref{fig:fibrepitch}.  In all three scenarios in this principle axis of muon momentum, smearing of the image was observed.  This is an inherent effect in the reconstruction of the interaction position in the principle axis associated with small angle (of several milliradians) scattering.    Cauchy-Lorentzian fits to the reconstructed $\lambda$ distributions yielded FWHM values of 5.70\,cm, 7.35\,cm and 12.15\,cm for the 1\,mm, 2\,mm and 4\,mm pitches respectively.

Figure~\ref{fig:fibrepitch} also highlights the typical differences observed in the reconstructed $\lambda$ values.  Although similar FWHM values were observed for the two smaller pitch fibres, the average $\lambda$ values reconstructed differed by approximately 25\%.  However, the degree of material discrimination provided by the 2\,mm pitch was still sufficient for this proof-of-principle prototype.  The suppression of the reconstructed $\lambda$ values, and consequently, the material discrimination ability with the 4\,mm fibre, are again highlighted.  These factors rendered the 4\,mm fibre pitch impractical for the location and discrimination of materials necessary for this application.

In addition to the differences in the reconstructed $\lambda$ values (vital for the discrimination of different materials) and the extent of the smearing in the z direction, the choice of fibre pitch impacted on the cost of read out electronics and complexity of construction.   Factoring in the anticipated increase in cost and the higher degree of intricacy arising from the 1\,mm pitch, the decision was made to construct the prototype system using 2\,mm-pitch fibres.  All studies presented in the following sections were therefore performed using only this pitch.

\subsection{Module Separations}\label{sub:spacings}
Similar to the choice of fibre pitch, there were several factors which served to influence the choice of z positions for the four detector modules: the reconstructed image resolution, the expected muon flux, the structural rigidity of the vertical support frame, and the size of the assay volume necessary to accommodate proposed test objects.   With a total separation denoted $\Delta$z$_{\mathrm{outer}}$, of 900\,mm between the outermost detector modules imposed by the external support frame to negate any destabilising effects, and a proposal to image a small-scale waste barrel of 250\,mm height in the future, all studies performed were subject to this $\Delta$z$_{\mathrm{outer}}$ constraint and a fixed separation between the two innermost modules denoted $\Delta$z$_{\mathrm{inner}}$, of 400\,mm to allow for mechanical support.

With the limited acceptance of the proposed prototype system, it was important to maximise the cosmic-ray muon flux through all four modules.  This was particularly vital for the application in question, though in reality, the required muon exposure durations for this small acceptance prototype would be significantly longer than required for a proposed full-scale system.  The expected muon flux scaled, roughly, with the acceptance as $1/\Delta$z$_{\mathrm{outer}}^2$.  Therefore, to increase the number of muons detected (and to minimise the data collection duration required) it was important that this distance was kept as small as possible.  A reduction in this value of 100\,mm from the maximum permitted separation corresponded to an increase in statistics of approximately 25\%.

In principle however, the image resolution obtained by the system should be enhanced by increasing the size of $\Delta$z$_{\mathrm{top}}$ and $\Delta$z$_{\mathrm{bottom}}$, \ie the separations between the topmost and bottommost detector modules.

Negligible differences were observed in the image resolution obtained from several different combinations of $\Delta$z$_{\mathrm{outer}}$,  $\Delta$z$_{\mathrm{top}}$ and $\Delta$z$_{\mathrm{bottom}}$ given the small active area (and hence the small angular acceptance) of the system.  These studies will however be essential for the development of any larger scale detector configuration where this effect will be significant. 

\begin{figure*}[t] 
\centering 
\includegraphics[width=0.66\columnwidth,keepaspectratio]{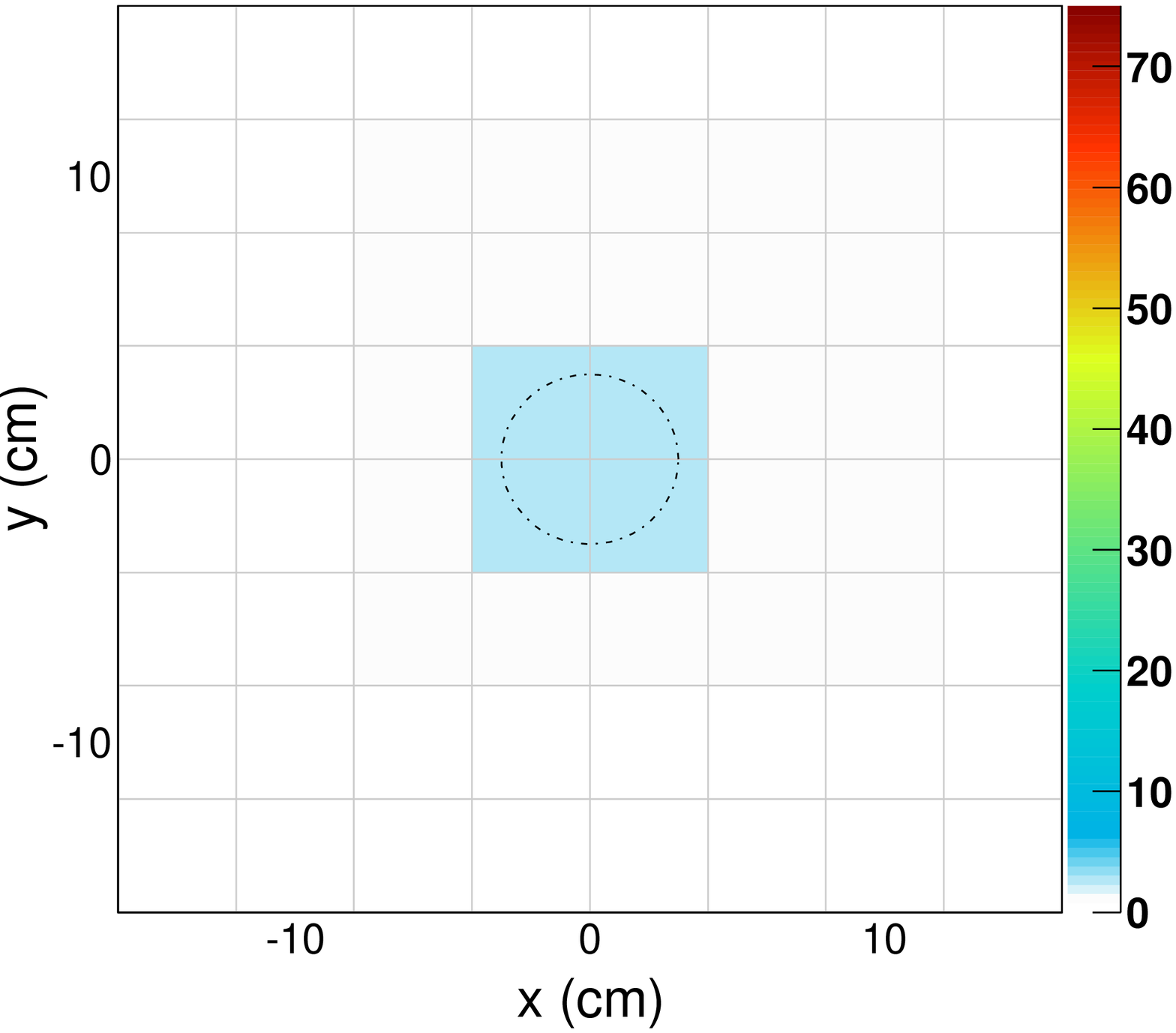}
\includegraphics[width=0.66\columnwidth,keepaspectratio]{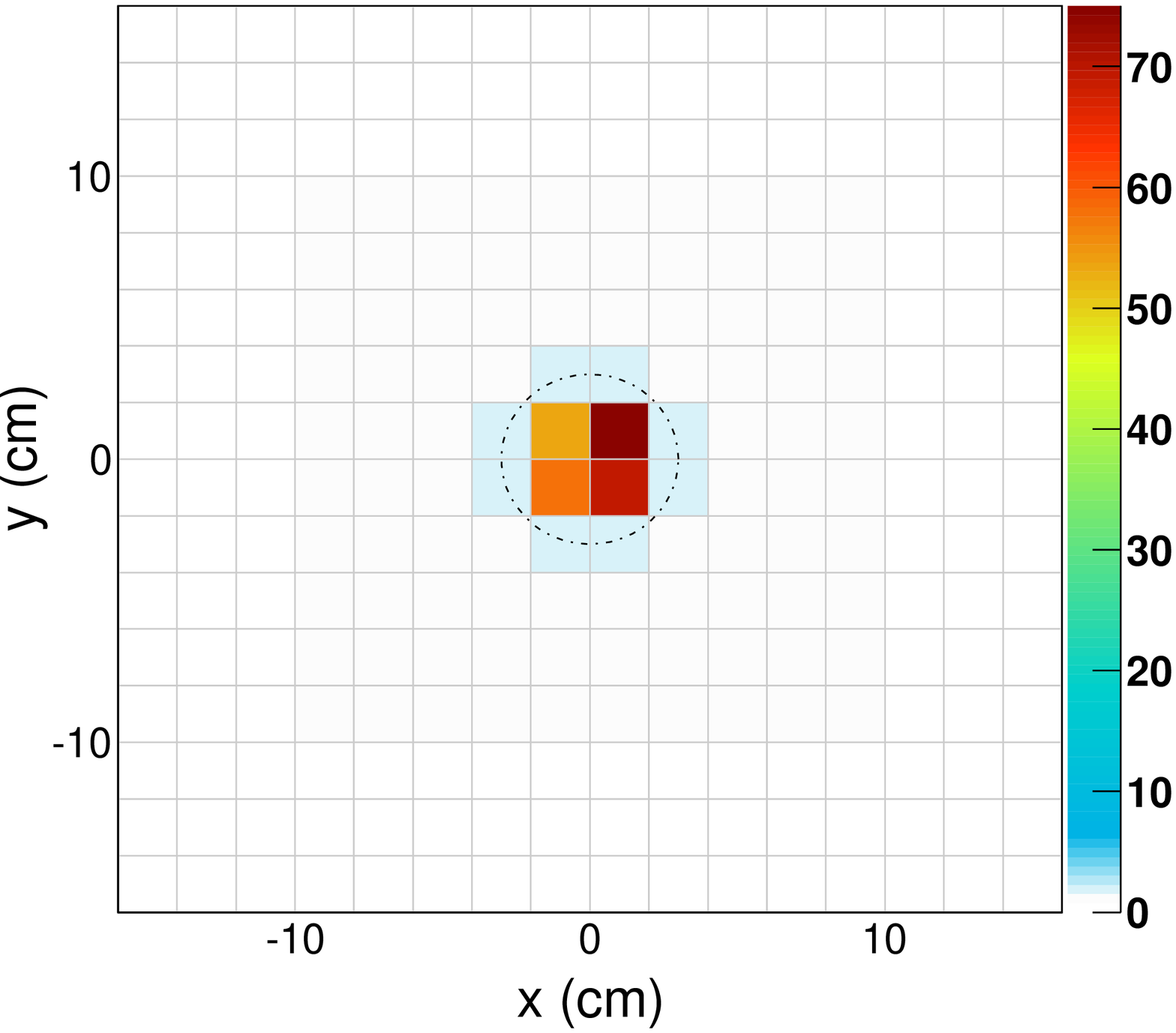}
\includegraphics[width=0.66\columnwidth,keepaspectratio]{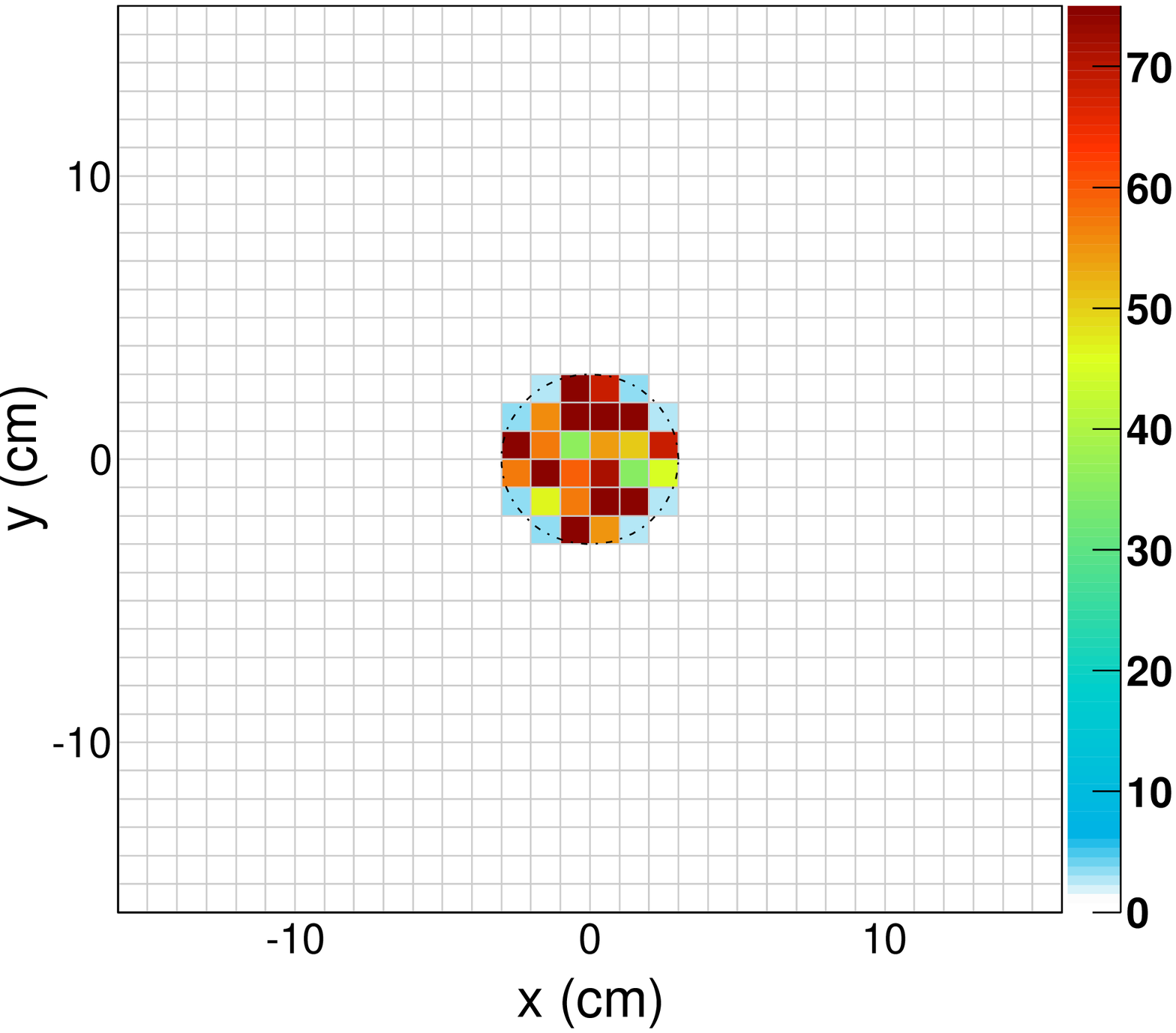}
\caption{Comparison between the reconstructed images in the central 10\,mm slice in the xy plane using 40\,mm (left), 20\,mm (centre) and 10\,mm (right) voxel dimensions in the x and y directions.  The voxel grids are shown on each image along with the outline of the uranium sphere.  As anticipated, greater definition and material characterisation in the reconstructed image was observed with smaller voxel dimensions.}
\label{fig:VoxSizeXY}
\end{figure*}

The results presented in the following studies used spacings of 900\,mm, 400\,mm, 250\,mm and 250\,mm for $\Delta$z$_{\mathrm{outer}}$, $\Delta$z$_{\mathrm{inner}}$, $\Delta$z$_{\mathrm{top}}$ and $\Delta$z$_{\mathrm{bottom}}$ respectively.  

\subsection{Voxel Size Studies}
Factoring in the 80\% detection efficiency per layer obtained from simulation and studies performed on a constructed module~\cite{VCI}, approximately 1000 muons per day were expected to scatter within the chosen 300\,mm dimension cubic assay volume and pass all relevant selection criteria.   Figure~\ref{fig:VoxSizeXY} shows the images obtained from a simulation of a uranium sphere of 30\,mm radius at the centre of this volume.  These images used 10\,mm, 20\,mm and 40\,mm voxels (in both the x and y directions, and 10\,mm in z).  All images were reconstructed using the same muon exposure time. 

In the case of the largest voxels, a square of seemingly low-$Z$ material composition was reconstructed.  This was due to the fact that the voxels contained mixed contributions from both uranium and air.  Here, the latter contribution served to dilute the effective scattering density reconstructed within the voxel.  The dimensions of this image were over twice the size of the simulated object.   In contrast, with 10\,mm voxels, a high-$Z$ object was identified of comparable dimensions to that simulated.  However, (and recalling that the z dimensions of the voxels are constant across all three scenarios) this defined image would require approximately 16-times longer to resolve.  All studies presented in the remainder of this work were performed used cubic voxels of 10\,mm dimension.   More exposure time comparisons are presented in Section~\ref{sub:matdiscrim}.

\begin{figure}[hb] 
\centering 
\includegraphics[width=1.0\columnwidth,keepaspectratio]{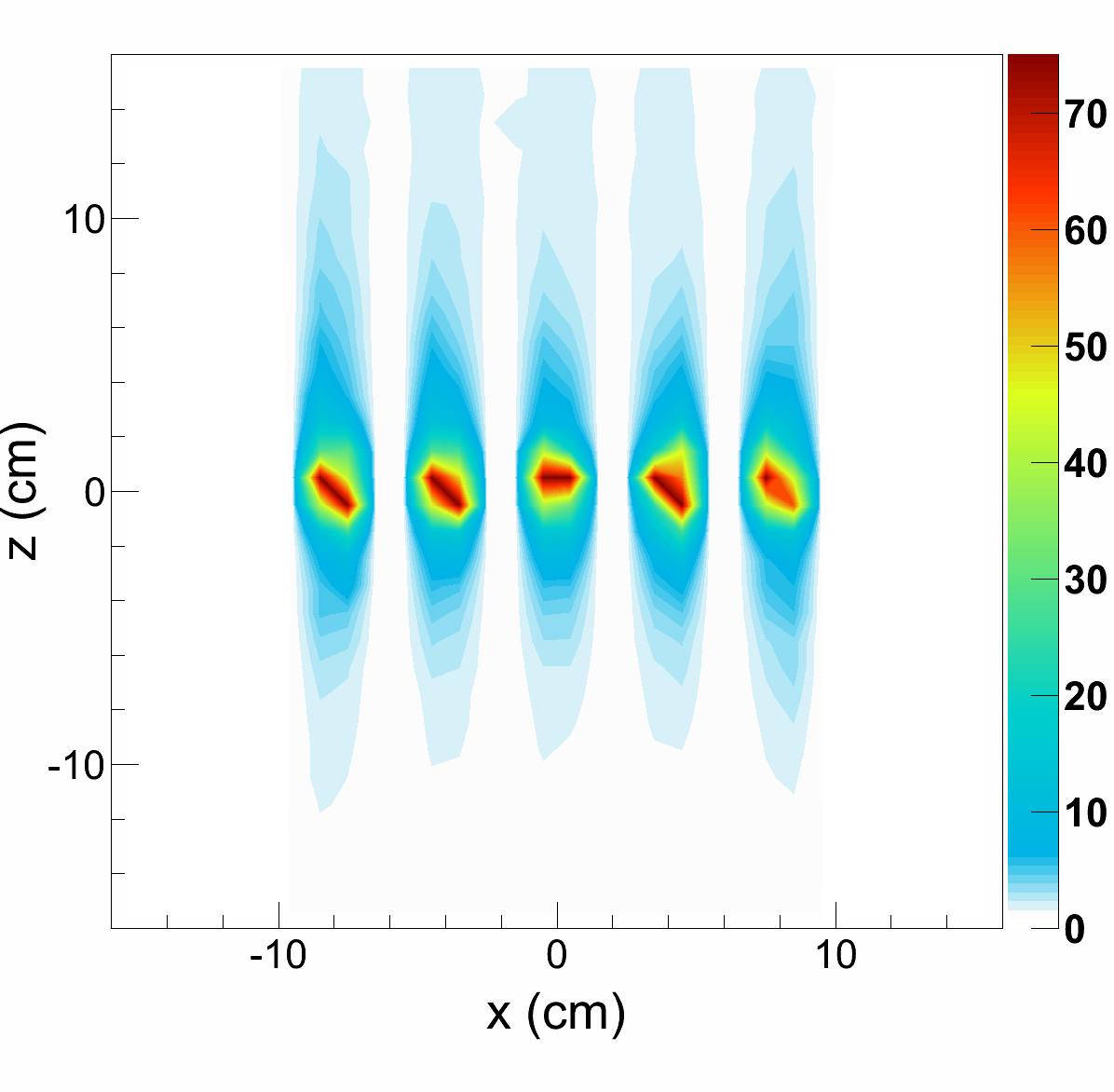}
\caption{Image highlighting the increasing effect of the z-smearing at the edges of the detector acceptance.  Simulated are five cubic blocks of uranium of 40\,mm dimension within an air matrix at the centre of the assay volume.  The FWHM values range from 12\,mm for the central block to approximately 20\,mm at the edges of the acceptance.  A contoured plotting style was chosen to best highlight this effect.  The quoted FWHM values differ to the values quoted previously due to the smaller piece of uranium under interrogation causing less Coulomb scattering and differing iteration requirements.}
\label{fig:acceptancesmear}
\end{figure}
\subsection{Acceptance \& Smearing Effects}
Simulated data were analysed to assess the material reconstruction ability and the extent of the z-smearing at the edges of the prototype acceptance.  This effect, which was observed in the studies presented in Section~\ref{sub:pitch}, was artificially exaggerated in this work as a result of the limited angular acceptance provided by the 256\,mm\,x\,256\,mm active area and 900\,mm height of the system, which equated to approximately 80\,msr in solid angle coverage.  

\begin{figure*}[t] 
\centering 
\includegraphics[width=0.495\columnwidth,keepaspectratio]{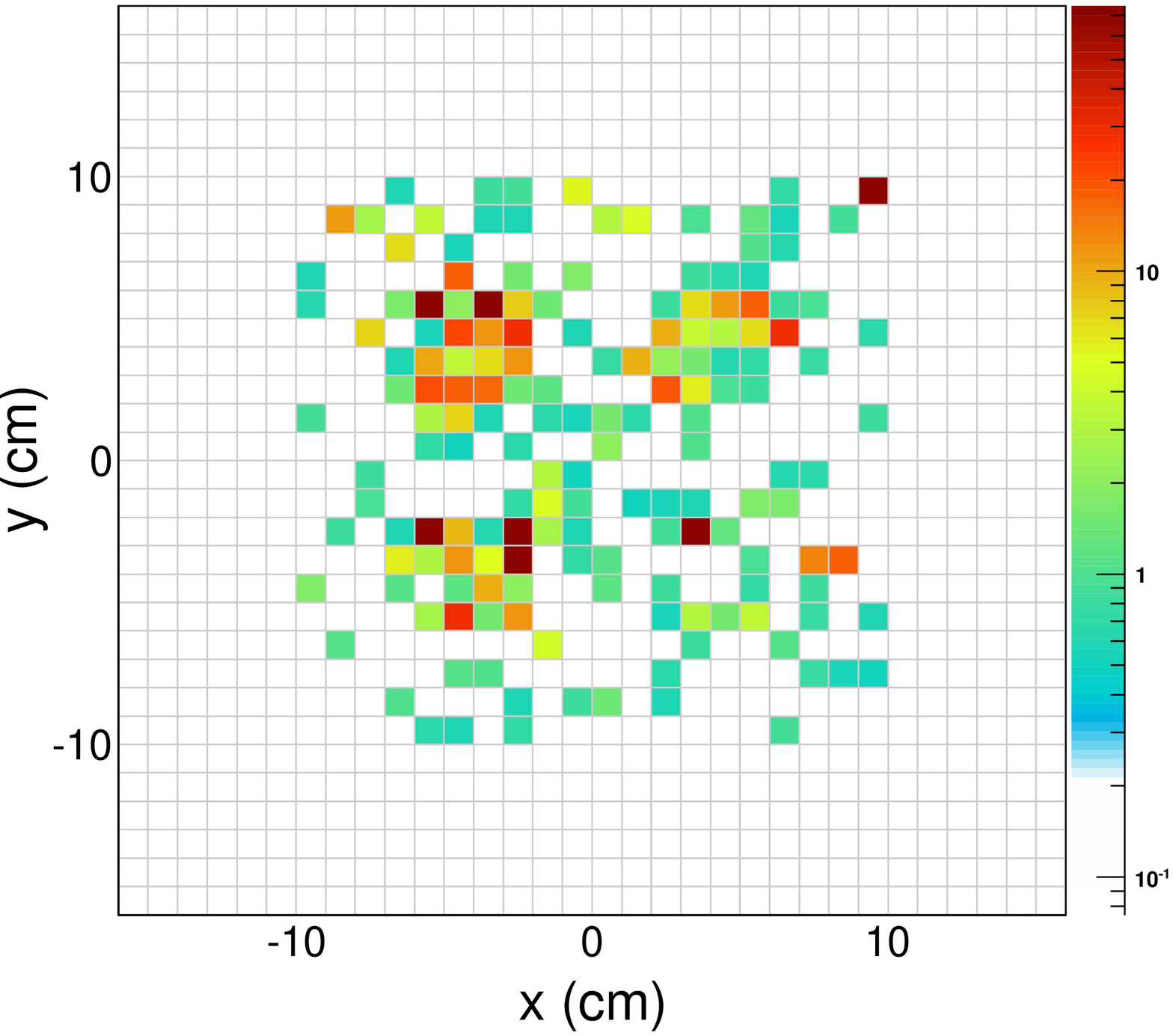}
\includegraphics[width=0.495\columnwidth,keepaspectratio]{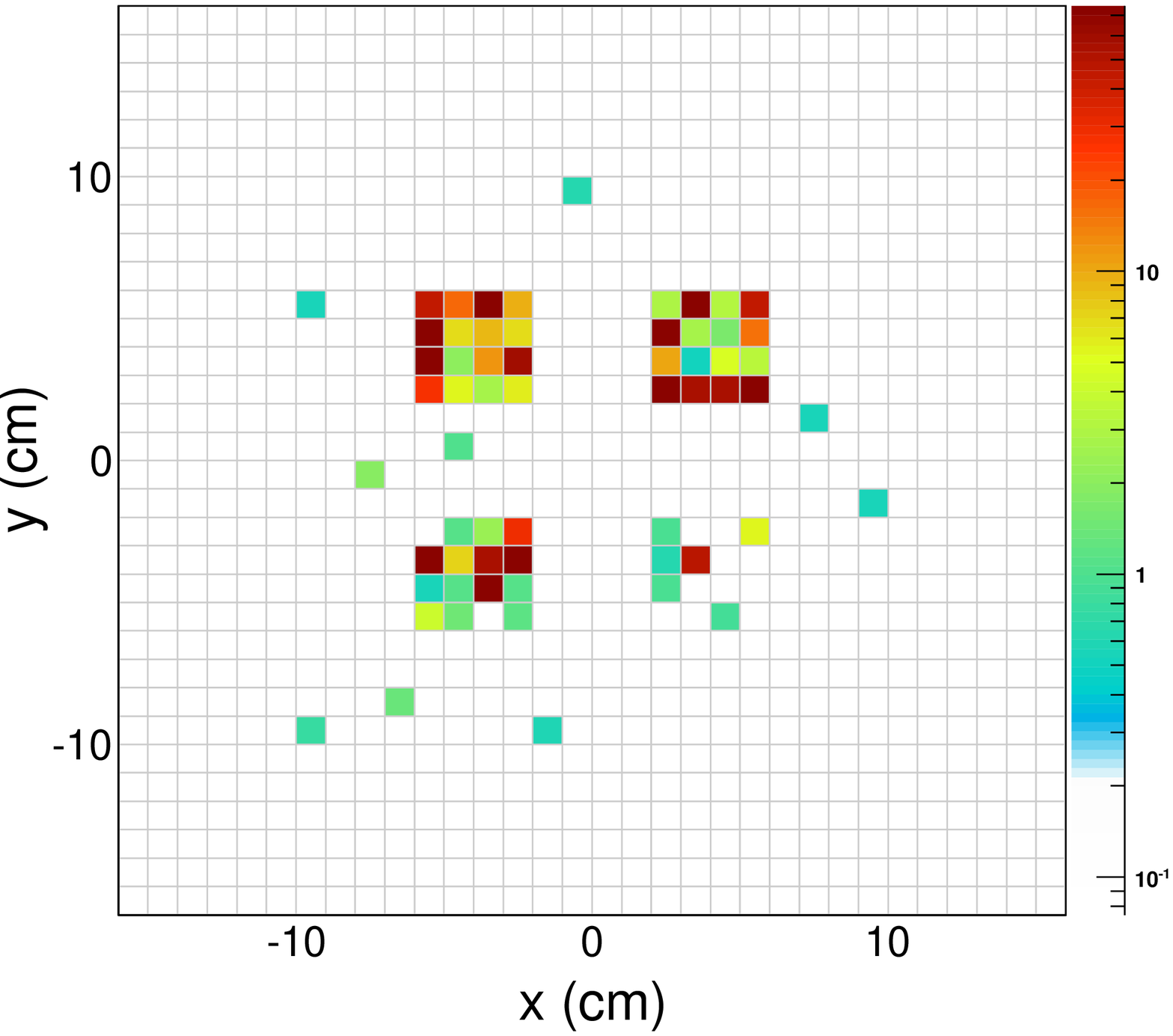}
\includegraphics[width=0.495\columnwidth,keepaspectratio]{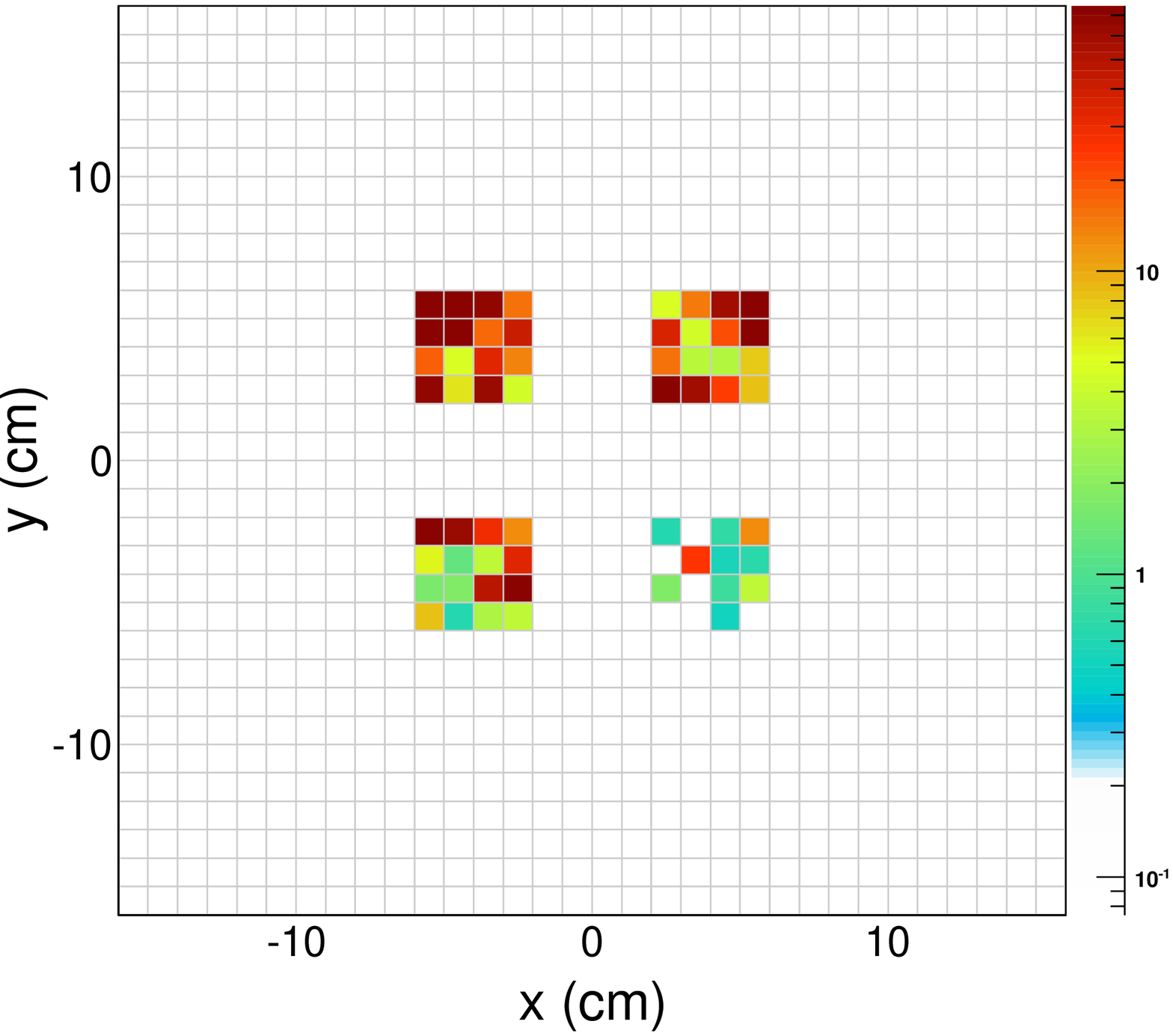}
\includegraphics[width=0.495\columnwidth,keepaspectratio]{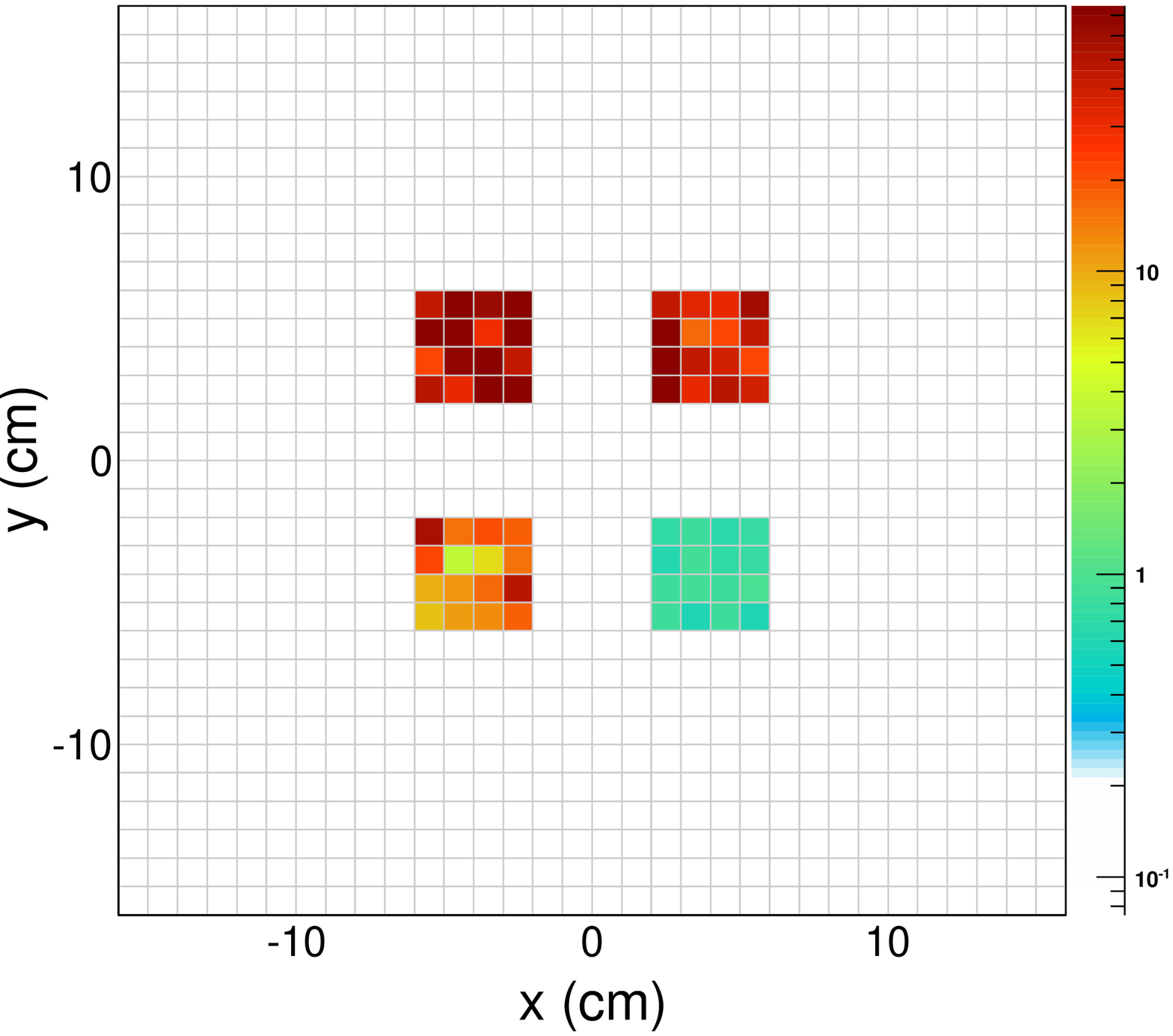}
\caption{Images reconstructed from 1 day (far-left), 1 week (middle-left), 1 month (middle-right) and 1 year (far-right) of muon exposure using the small-scale prototype, highlighting the expected material discrimination from the system.  For each duration, 40\,mm cubic blocks of (clockwise from top-left) uranium, uranium oxide, concrete and iron were simulated within an air matrix.  The symmetric orientation of these cubes ensured that similar muon flux was experienced by each object.   Here, a logarithmic colour scale has been chosen to best highlight the discrimination between the four different materials.  Each figure is described in detail in the text.} 
\label{fig:materialdisc}
\end{figure*}
Five 20\,mm cubes of uranium were simulated in the xz-plane at the centre of the assay volume with 20\,mm separations in the x direction.   Figure~\ref{fig:acceptancesmear} shows the reconstructed image from the central 20\,mm slice in the xz-plane through the centre of the assay volume which these blocks fully occupied.  The bulk of the uranium was detected within the 20\,mm region in the z-direction in which they were simulated, and the extent of the smearing in this direction appeared uniform across the acceptance for medium to high values of $\lambda$.  At low values ($\lambda\,<$ 10\,mrad$^2$\,cm$^{-1}$) an expected increase in the effect of this smearing was observed at the limits of the acceptance.  The FWHM values from Cauchy-Lorentzian fits to these z-distributions increasing from 12\,mm for the central cube to values in the region of 20\,mm for those at the edges.  This inherent effect is expected to decrease significantly with an increase in angular acceptance.

The FWHM values obtained from Figure~\ref{fig:acceptancesmear} are smaller than those quoted previously due to the reduction in size of the uranium blocks simulated and the resultant reduction in Coulomb scattering experienced by the muon. 

\subsection{Material Discrimination}\label{sub:matdiscrim}
The most important requirement for this application was the identification and characterisation of high-$Z$ waste products stored within more dense material mixtures, predominantly, but not confined to, concrete-filled stainless steel containers.   It was important therefore that any constructed system could discern between these high-$Z$ materials and objects comprised of other low- or medium-$Z$ materials (\eg concrete, iron \etc) within a reasonable timescale.   To this end, a configuration of 40\,mm cubes of four different materials, namely uranium, uranium oxide, iron and concrete, was simulated in the central xy plane of the assay volume.  These materials had respective densities of 18.95\,g\,cm$^{-3}$, 10.96\,g\,cm$^{-3}$, 7.87\,g\,cm$^{-3}$ and 2.30\,g\,cm$^{-3}$.  All four blocks were oriented such that each experienced a similar muon flux.  An air matrix was used for this study.   The effect of imaging materials within a concrete matrix is studied in Section~\ref{sub:concrete}.

Figure~\ref{fig:materialdisc} shows the images reconstructed in a 10\,mm slice in the xy plane through the centre of the simulated cubes for four simulated real-time durations ranging from one day of data collection to an idealised case of one year of cosmic-ray muon exposure.  These timescales for the small-scale prototype are in no way indicative of anticipated data collection durations in an industrial environment with a full-scale system, and served only to highlight the improved material discrimination observed with increased statistics.

After only one day of simulated muon exposure, there were indications that the three denser objects were present, though the high extent of noise which existed within the reconstructed image (and also throughout the assay volume) did not allow any material identification to take place.  With the exposure time increased to one week, the dimensions of these three blocks became clear, though all three could be interpreted as being of similar material composition or density due to the non-uniformity of $\lambda$ values reconstructed within each block.  With reduced noise in the image, there were clearer indications that the fourth object, the concrete cube, was also present though this appeared smaller than that simulated.  Also, no firm statement could be made as to the relative density of this object.   After one full month of data collection, all four objects were clearly observed and material discrimination of low (air and concrete), medium (uranium oxide and iron) and high-$Z$ (uranium) materials could be obtained.  The final, idealised timescale revealed the optimal level of discrimination achievable with this prototype MT system.  All four materials were distinguished from each other and the $\lambda$ fluctuations within each material observed for the shorter timescales were significantly reduced.

\subsection{Concrete Matrix Studies}\label{sub:concrete}
\begin{figure}[hb] 
\centering 
\includegraphics[width=0.495\columnwidth,keepaspectratio]{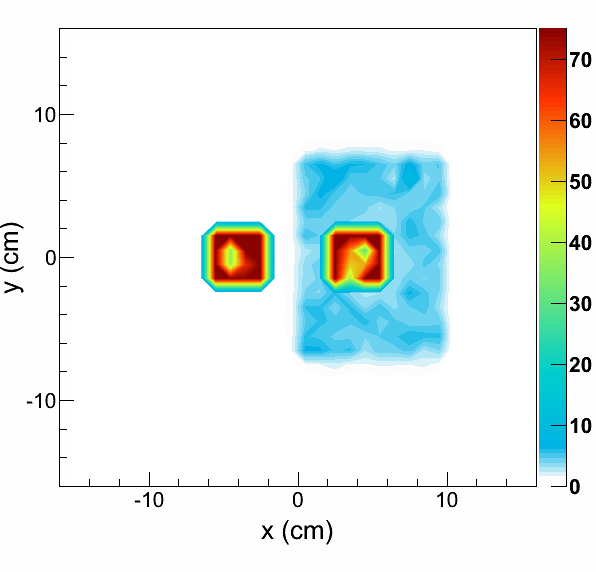}
\includegraphics[width=0.495\columnwidth,keepaspectratio]{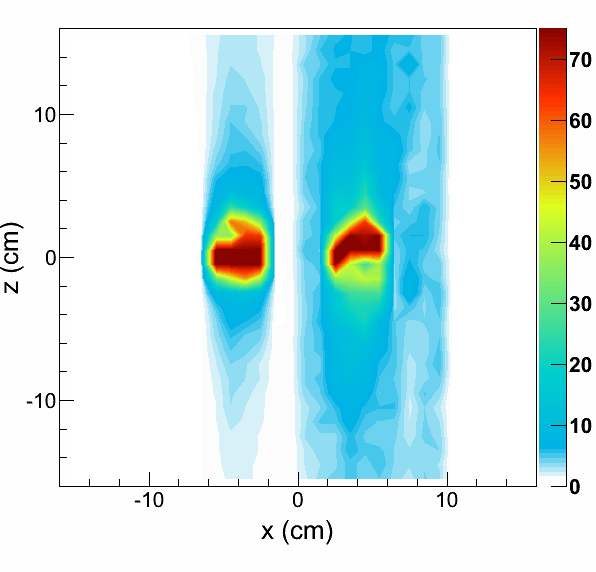}
\caption{Tomograms of 10\,mm width reconstructed in the xy (left) and xz (right) planes from a test setup containing two 40\,mm cubes of uranium, one simulated in air (negative x direction) and one in concrete (positive x direction).  No discernible difference is observed in the spread of converged $\lambda$ values within the two blocks.  A slight increase, as expected, in the smearing effect was observed within the concrete matrix due to the increase in Coulomb scattering experienced by the muons in the concrete.}
\label{fig:Concrete}
\end{figure}

All studies presented previously have reconstructed objects simulated within an air matrix.   For the proposed industrial application, the identification and characterisation of high-$Z$ material within a denser, concrete matrix was essential.  Higher levels of Coulomb scattering were expected compared with the air scenario.  It was initially unclear whether this would result in a significant increase in smearing or a dilution in the reconstructed $\lambda$ values, rendering the necessary characterisation difficult.  To determine this, a test case consisting of two 40\,mm cubes of uranium centred on ($\pm$40\,mm,\,0,\,0) were simulated with an air (concrete) matrix in the negative (positive) x-direction.   Figure~\ref{fig:Concrete} shows the image reconstructed from a 10\,mm slice in the xy and yz planes through the centre of both blocks.   The smearing in the z-direction was accentuated in the concrete matrix as highlighted by the increased FWHM values observed in Figure~\ref{fig:MomComparison}.  However, the $\lambda$ ranges of the uranium block in both cases were compatible, suggesting the material discrimination was not significantly adversely affected by this denser matrix.

\subsection{Momentum Dependence}\label{sub:momenta}
All results presented in the previous subsections have assumed that no muon momentum measurement was available.  As such, each muon was assigned a momentum of 3.35\,GeV\,c$^{-1}$, equal to $p_{\mathrm{mean}}$, the mean of the sea-level cosmic-ray muon spectra.   In reality, the determination of $\lambda$ within the MLEM algorithm exhibited a dependence on the individual momentum of each muon $p_i$ via the scaling factor $p_{i}/p_{\mathrm{mean}}$.  This had the effect of suppressing the contribution to the overall $\lambda$ determination from low-momentum muons, which experienced more individual scatters and thus could scatter through uncommonly large angles.  This also had the detrimental impact of creating erroneous values of $\lambda$ resulting from the large, though less probable, tails of the Cauchy-Lorentzian distributed scattering angles.   The studies shown in the previous subsections, which neglected momentum information, suggested that the determination of the muon momentum was not crucial to the identification and characterisation of different materials, even within a concrete matrix.  

\begin{figure}[t] 
\centering 
\includegraphics[width=\columnwidth,keepaspectratio]{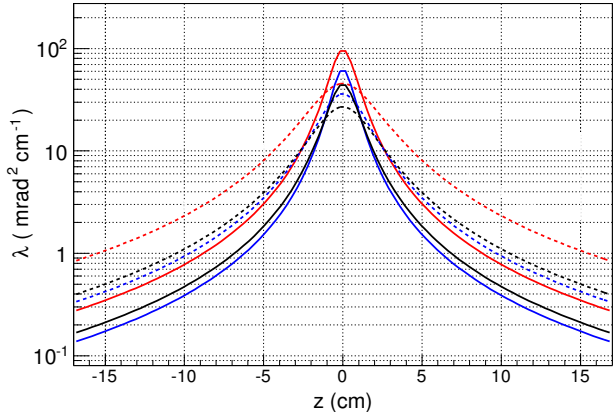}
\caption{Comparison between the Cauchy-Lorentzian fits to the z-smearing of reconstructed uranium blocks in an air (solid line) and concrete (dashed line) matrices with perfect momentum reconstruction (blue), 20\% resolution in momentum reconstruction (black) and no momentum information (red).  For the latter case, all muons are treated as having the mean momentum of the cosmic-ray muon distribution.  All distributions have been corrected for minor millimetre-order shifts in z for ease of comparison.  These fits are described in the text.}
\label{fig:MomComparison}
\end{figure}

Dedicated studies were performed to assess the potential improvement gained with the knowledge of individual momenta.  Three scenarios were compared:  perfect measurement of the momentum, a realistic 20\% uncertainty in the momentum measurement, and the standard case of no momentum measurement whereby each muon was assigned a value of $p_{\mathrm{mean}}$.   

The same test configuration from Section~\ref{sub:concrete} was simulated.  Figure~\ref{fig:MomComparison} shows Cauchy-Lorentzian fits to the reconstructed $\lambda$ distributions in the z direction from tomograms through the centre of both blocks for each momentum scenario.  The results reveal the similar distributions obtained from the two scenarios which benefit from momentum measurements.   With no measure of momentum for each individual muon, the reconstructed $\lambda$ values were found to be systematically higher in both the air and concrete matrices.    The larger FWHM values, which were up to 40\% larger in the concrete case, indicated a reduced discrimination between the uranium and background material.   A similar reasoning applied to the comparison between air and concrete matrices that has been discussed previously.

Despite the improvement in the smearing and (close to) absolute determination of $\lambda$, the results obtained here, and earlier in this work, show that the ability to identify and discern low, medium and high-$Z$ materials was not significantly affected by the lack of momentum information.  This finding, and the added cost to a constructed system in obtaining a momentum measurement was used to conclude that no measurement was necessary for the purposes of this proof-of-principle system.

\begin{figure}[t] 
\centering 
\includegraphics[width=0.7\columnwidth,keepaspectratio]{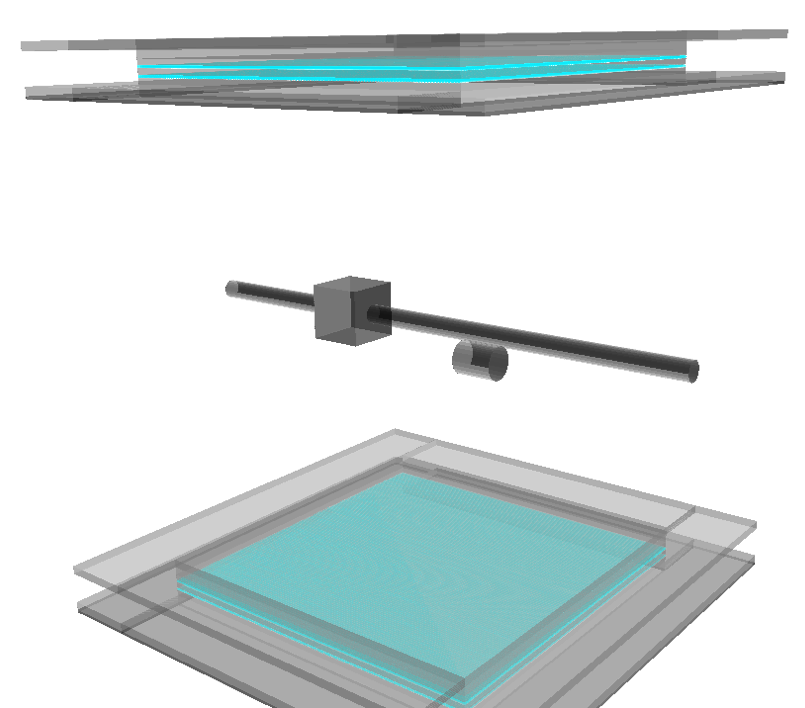}
\caption{GEANT4 simulation of the active components of the prototype detector system showing the configuration of lead, uranium and stainless steel within the central region of the assay volume.  Not pictured are the outermost detector modules and support structures as described in the text.  The \roha (translucent grey), orthogonal layers of fibres (blue) and aluminium baseplate are all shown.}
\label{fig:SetupSimulation}
\end{figure}
\section{Simulation Verification With Experimental Data}\label{sec:TestSetup}
A prototype MT system was constructed based around the optimal specifications outlined in the previous section.  Details regarding the design and fabrication of this system are found in Ref.~\cite{VCI}.  Data were collected over a period of 24\,weeks using this system with a test configuration of objects within the assay volume.  This configuration consisted of a 40\,mm cube of lead surrounding a 12\,mm diameter stainless steel rod from which a cylinder of uranium metal 30\,mm-long and 20\,mm in diameter was suspended.   The GEANT4 simulation of this setup is visualised in Figure~\ref{fig:SetupSimulation}.  The images reconstructed from experimental data were compared with those obtained from detailed simulation of this configuration for the same muon exposure duration in order to validate the studies performed previously.

\begin{figure*}[t] 
\centering 
\includegraphics[width=\columnwidth,keepaspectratio]{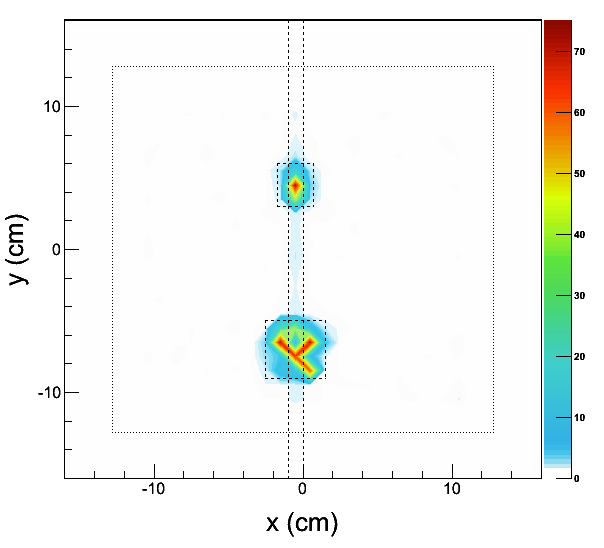}
\includegraphics[width=\columnwidth,keepaspectratio]{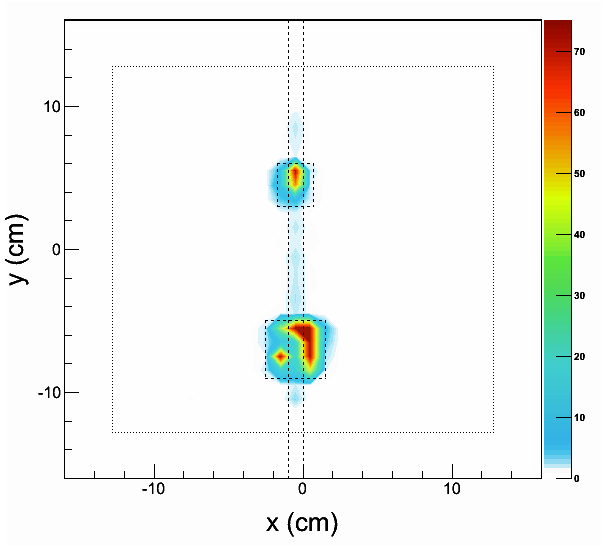}
\caption{Comparison between images reconstructed from 24\,weeks of exposure to cosmic-ray muons for GEANT4 simulation (left) and experimental data (right).  Shown is a 10\,mm slice in the $xy$-plane \ie parallel to the detector modules. The dashed lines provide an estimate of the location and dimensions of the test objects in the assay volume.  The active area of the detector system is also indicated by the fine-dotted square.  The uranium (top), lead (bottom) and stainless-steel bar are all clearly visible within the air matrix ($\lambda <$ 1\,mrad$^{2}$\,cm$^{-1}$).  Here, the colour scale denotes the most-likely $\lambda$ value in each voxel reconstructed by the MLEM algorithm.  Excellent agreement is observed between the results from data and simulation.  This is described in detail in the text.}
\label{fig:XYcomparison}
\end{figure*}

\begin{figure*}[t] 
\centering 
\includegraphics[width=\columnwidth,keepaspectratio]{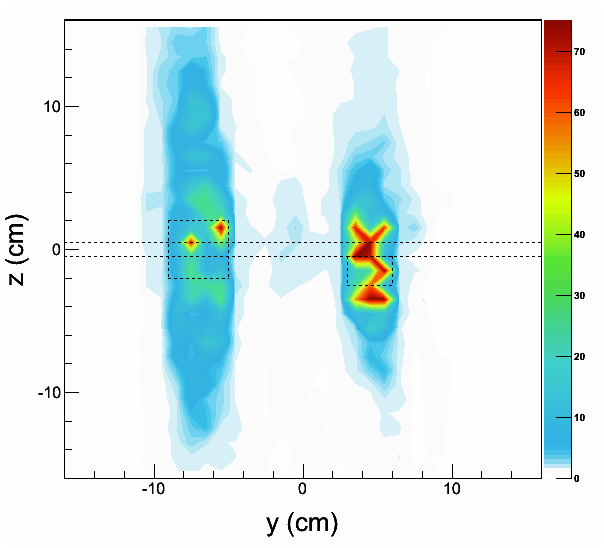}
\includegraphics[width=\columnwidth,keepaspectratio]{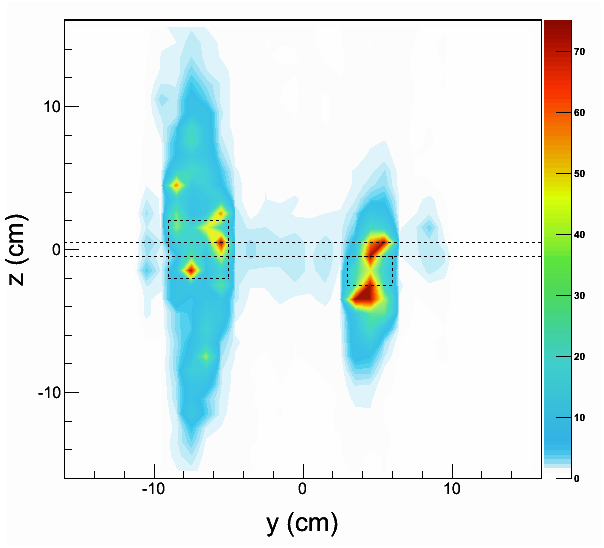}
\caption{Comparison between images reconstructed from 24\,weeks of exposure to cosmic-ray muons for GEANT4 simulation (left) and experimental data (right).  Shown is a 10\,mm slice in the $yz$-plane \ie side-on to the detector modules. The dashed lines provide an estimate of the location and dimensions of the test objects in the assay volume.   The lead (left), uranium (right) and stainless-steel bar are all clearly visible within the air matrix ($\lambda <$ 1\,mrad$^{2}$\,cm$^{-1}$).  Here, the colour scale denotes the most-likely $\lambda$ value in each voxel reconstructed by the MLEM algorithm.  Excellent agreement is observed between the results from data and simulation.  This is described in detail in the text.}
\label{fig:YZcomparison}
\end{figure*}

Results showing 10\,mm tomograms in the xy and yz planes through the length of the stainless steel bar are shown in Figures~\ref{fig:XYcomparison} and \ref{fig:YZcomparison} respectively.   Initial results detailing the experimental data collection and event selection criteria were shown previously in comparison with simulated results in the xy plane only in Ref.~\cite{VCI}.   These results confirmed the validity of the simulations discussed in the previous sections.   The updated experimental results shown in this work show excellent agreement between the simulated results and those reconstructed from data collected by the prototype detector for the same 24\,week duration.   As expected, in both the simulation and experimental data, the three objects were observed within their expected boundaries in the xy plane.  This again highlighted the precision image resolution within the directions orthogonal to the principle axis of muon momentum.  Clear discrimination was observed between the two high-$Z$ materials, the stainless steel bar and the surrounding air matrix.     From this tomogram, material discrimination between the two high-$Z$ objects was inconclusive.   The uranium however, was actually situated below the level of the stainless steel bar and as such it did not occupy the voxels shown in Figure~\ref{fig:XYcomparison}.  Instead, the effects of the z-smearing are observed here.

In contrast, the tomograms in the yz plane from the simulated and experimental data, revealed a distinct region of $\lambda$ values above 60\,mrad$^{2}$\,cm$^{-1}$ consistent with the uranium location.  The lead volume was found to possess a scattering density in the region of 20\,mrad$^{2}$\,cm$^{-1}$ with minor fluctuations, possibly as a consequence of the lack of momentum information and/or increased multiple scattering due to the thickness of the high-$Z$ block.    These values are in agreement with those expected for these materials~\cite{Schultz04}.   This tomogram also revealed the similarity in the extents to which the two sets of data were affected by the smearing in the z direction.  The simulation, however, appeared to slightly overestimate this effect by approximately 10\%.

A possible reason for the non-uniformity within the expected object dimensions was the incomplete voxel coverage, \ie for the uranium cylinder, the voxels at the edges of this piece also contained contributions from air, which acted to dilute the signal observed.   The reconstruction of the steel bar was also clearer in the experimental data.  This was likely attributed to better voxel coverage in this region compared with the simulation and/or potential density differences for this material in both scenarios.

\section{Simulation of a Small-scale Waste Canister}\label{sec:canister}
To further confirm the feasibility of the constructed prototype system in the non-destructive assay of legacy waste containers, and indeed the imaging and simulation software developed for this purpose, a further simulation using the small-scale prototype was proposed.  For this, a small stainless-steel waste canister, filled with concrete and pieces of uranium and uranium oxide, was simulated.  The canister measured 175\,mm in diameter and 250\,mm in height.  This was approximately one fifth of the anticipated scale of an industrial waste container.  The canister was simulated with an exaggerated wall thickness of 12\,mm aimed to represent a worst case scenario for the real world application.  Inside the canister, two 40\,mm cubes, one of uranium and one of uranium oxide, were centred at $-$40\,mm and +40\,mm respectively in the x direction.   These materials represented a potential scenario of what could be expected in an industrial environment and aimed to assess the ability of this system to characterise the two nuclear materials present within this matrix.

\begin{figure*}[t] 
\centering 
\includegraphics[width=\columnwidth,keepaspectratio]{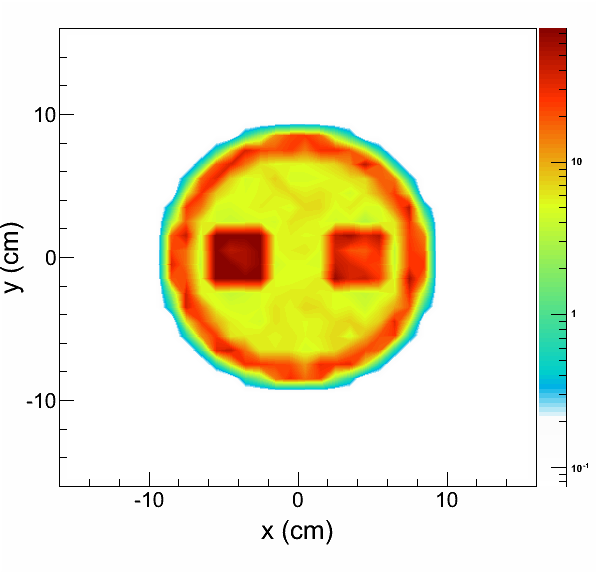}
\includegraphics[width=\columnwidth,keepaspectratio]{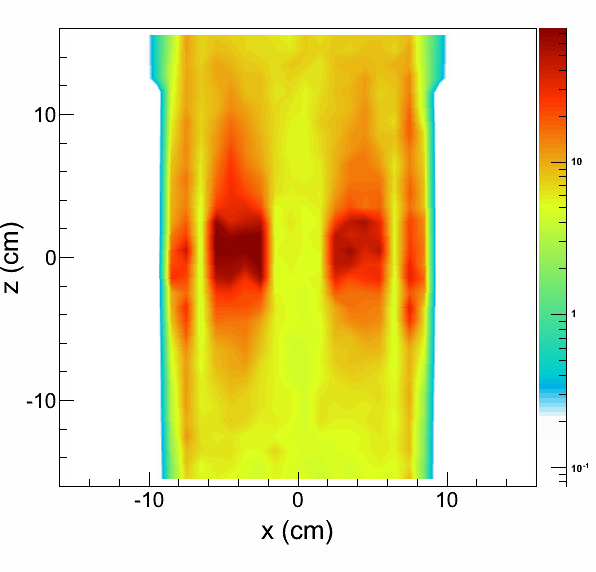}
\caption{Images reconstructed from a simulation of a small concrete-filled, stainless-steel canister using the small-scale prototype.  This canister contained two cubes, 40\,mm in dimension, of high-$Z$ materials: uranium (shown on the left-hand side within the barrel) and uranium oxide (right).  These are anticipated assay materials within the industrial waste containers.   Shown in the left (right) image is a 10\,mm tomogram in the $xy$ ($xz$) plane.  In both, the contours of the barrel and the two high-$Z$ materials within are clearly observed from several months of simulated exposure to cosmic-ray muons.   }
\label{fig:barrelimages}
\end{figure*}

Figure~\ref{fig:barrelimages} shows a 10\,mm tomogram in the central regions of both the xy and xz planes.   These have been reconstructed from several months of simulated muon exposure using the small scale MT system.   In the xy slice, the circular contour of the container was clearly observed against the air background.  More importantly, the two high-$Z$ materials of interest are both clearly visible within the concrete matrix of the canister.   The uranium was reconstructed with $\lambda$ values of over 60\,mrad$^{2}\,$cm$^{-1}$, whereas the less dense oxide showed scattering densities in the range 30$-$60\,mrad$^{2}\,$cm$^{-1}$.   The xz tomogram highlighted once again the limited acceptance of this prototype system which gave rise to smearing in the z direction.  Despite this effect, the vertical dimension of the canister could be inferred from the reconstruction of the stainless-steel casing which was visible within the simulated z range of [-12.5,12.5].  In the xy plane, this circular casing was clearly observed, and discriminated from the concrete, with lambda values in the region of 30\,mrad$^{2}\,$cm$^{-1}$.  The stainless-steel outer casing exhibited artificially-larger $\lambda$ values due to the extended amount of medium-$Z$ material in the vertical path of the muon. 
 
\section{Summary}
The feasibility of using a prototype muon tomography system, consisting of four scintillating-fibre tracker modules, for the detection and characterisation of high-$Z$ materials contained within nuclear waste containers has been confirmed via dedicated GEANT4 simulation studies.  These studies, performed prior to the construction of a fully-operational detector system directly influenced the selection of optimal fibre pitches and intermodule separations for experimental data collection.  First results from the project, presented in this work using the probabilistic Maximum Likelihood Expectation Maximisation image reconstruction technique, have confirmed the high-$Z$ material detection capabilities of this detector system and verify the initial, promising results from simulation studies.  Discrimination between low (air), medium (stainless steel) and high-$Z$ (lead and uranium) materials was observed with several months of cosmic-ray muon exposure.  Studies highlighted the high image resolution in the directions orthogonal to the muon motion, and the smearing observed in the parallel, z direction.  Initial simulation results showing the projected image resolution and material discrimination using this small-scale system with a small waste canister which mimicked the configuration expected in a realistic scenario, showed clear observations of two different high-$Z$ materials within this complex matrix.   It is foreseen that, with further development of this technology and software packages, a full-sized system will be employed in the future to assay the contents of legacy nuclear waste containers within the UK nuclear industry, and in doing so, significantly impact upon future storage policy by helping to mitigate the risks inherent with the long-term storage of nuclear waste.

\section*{Acknowledgements}
The authors gratefully acknowledge Sellafield Ltd., on behalf of the UK Nuclear Decommissioning Authority, for their continued funding of this project.


\bibliographystyle{plainnat}

\end{document}